# Stoichiometric Control and Optical Properties of BaTiO$_3$ Thin Films Grown by Hybrid MBE

Benazir Fazlioglu Yalcin[1], Albert Suceava[1], Tatiana Kuznetsova[1], Ke Wang[2], Venkatraman Gopalan[1], Roman Engel-Herbert[1,3]

[1] *Materials Science and Engineering, The Pennsylvania State University, University Park, PA, USA*

[2] *Materials Research Institute, The Pennsylvania State University, University Park, PA, 16802, USA*

[3] *Paul-Drude-Institute for Solid State Electronics, Hausvogteiplatz 5-7, 10117 Berlin, Germany*

*Correspondence should be addressed to: bmf5511@psu.edu and engel-herbert@pdi-berlin.de

Keywords: *BaTiO$_3$, thin films, hybrid molecular beam epitaxy, optical properties*

**Abstract**

BaTiO$_3$ is a technologically relevant material in the perovskite oxide class with above-room-temperature ferroelectricity and a very large electro-optical coefficient, making it highly suitable for emerging electronic and photonic devices. An easy, robust, straightforward, and scalable growth method is required to synthesize epitaxial BaTiO$_3$ thin films with sufficient control over the film's stoichiometry to achieve reproducible thin film properties. Here we report the growth of BaTiO$_3$ thin films by hybrid molecular beam epitaxy. A self-regulated growth window is identified using complementary information obtained from reflection high energy electron diffraction, the intrinsic film lattice parameter, film surface morphology, and scanning transmission electron microscopy. Subsequent optical characterization of the BaTiO$_3$ films by spectroscopic ellipsometry revealed refractive index and extinction coefficient values closely resembling those of stoichiometric bulk BaTiO$_3$ crystals for films grown inside the growth window. Even in the absence of a lattice parameter change of BaTiO$_3$ thin films, degradation of optical properties was observed, accompanied by the appearance of a wide optical absorption peak in the infrared spectrum, attributed to optical transitions involving defect states present. Therefore, the optical properties of BaTiO$_3$ can be utilized as a much finer and more straightforward probe to determine the stoichiometry level present in BaTiO$_3$ films.

# 1. Introduction

BaTiO$_3$ thin films have attracted tremendous attention in recent years due to their high dielectric constant as well as ferroelectric and optical properties.[1–9] In particular, this interest in BaTiO$_3$ has been motivated by its superior properties such as low dissipation factor[10], low leakage current[11], high breakdown fields[12], and large electro-optical coefficients[13–16]. The relatively close resemblance of its lattice parameter with conventional semiconductors exhibiting diamond structure, specifically Si and Ge, has allowed BaTiO$_3$ to be epitaxially integrated with these material platforms and maintain its functionality.[17–20] BaTiO$_3$ is a prototypical perovskite oxide in the ABO$_3$ form with a Curie temperature of ~120 °C. At room temperature, it is tetragonal with lattice parameters a=3.992 Å and c=4.036 Å, with ferroelectric polarization along the c-axis.[21] BaTiO$_3$-based materials are generally regarded as a very interesting research topic due to their attractiveness for many applications such as FerroFETs to overcome the von Neumann bottleneck[22], high-speed optical modulators exploiting the Pockels effect[1], and quantum computing.[23]

The BaTiO$_3$ thin films have been prepared by various synthesis methods, such as sputtering[24,25], metal-organic chemical vapor deposition (MOCVD) [26–28], ion beam deposition[29], pulsed laser deposition (PLD)[4,30–33], RF-magnetron sputtering[34], and molecular beam epitaxy (MBE) [3,16,18,35]. Although these conventional thin film growth techniques can produce high-quality oxide films, good compositional control requires precise and direct flux ratio adjustment of the source material, which has been proven challenging. In some conventionally employed thin film synthesis approaches, a direct and independent flux control is not feasible. Even in the case of supplying a stoichiometric ratio, the formation of intrinsic defects, such as Frenkel and Schottky defects, could arise when growing thin films under conditions far from equilibrium. The situation is further complicated by the epitaxial constraint imposed by the substrate, which could lead to strain relaxation and the incorporation of additional crystallographic defects.

In comparison to above mentioned thin film growth techniques, hybrid molecular beam epitaxy (hMBE), which has been demonstrated to provide superior control over cation stoichiometry in complex oxide thin films, offers the advantage to access a self-regulated growth window.[36–38] In contrast to conventional MBE, where all cations are supplied by thermal evaporation together with an oxidizing agent – most prominently molecular or atomic oxygen, or ozone – in hMBE, only one of the elements, typically the A-site cation, is supplied by thermal evaporation in an effusion cell, while the B-site cation is supplied in the form of a metal-organic precursor and oxygen is shuttled to the growth front within this precursor molecule as well.[38,39] The volatility of the metal-organic precursor enables self-regulated growth kinetics, opening up an adsorption-controlled growth regime in a window of different flux ratios of the thermally evaporated A-site cation and the metal-organic precursor.[38] $BaTiO_3$[8,9] and solid solutions of $(Ba,Sr)TiO_3$[40,41] have already been successfully grown by hMBE utilizing the metal-organic precursor titanium(IV) tetraisopropoxide (TTIP) as Ti source. A growth window was reported for $BaTiO_3$ grown on $GdScO_3$ (110) substrates, identified by constant out-of-plane lattice parameters of coherently strained $BaTiO_3$ films.[9] In another study, the growth window of hMBE-grown $BaTiO_3$ was determined from reflection high-energy electron diffraction (RHEED) and film surface morphologies.[8]

Here, we study the stoichiometric control of hMBE-grown $BaTiO_3$ thin films on $(LaAlO_3)_{0.3}(Sr_2AlTaO_6)_{0.7}$ (LSAT) by systematically varying the TTIP flux while keeping the Ba flux fixed. The existence of a growth window was established using RHEED images, film surface morphology, the film lattice parameter, and scanning transmission electron microscopy. Different than the previous work on hMBE-grown BaTiO$_3$, spectroscopic ellipsometry was performed in the spectral range from 375 nm to 1690 nm to extract the refractive indices and extinction coefficients of BaTiO$_3$ films. It is shown that for optimal growth conditions within the growth window bulk-like optical properties were obtained, while deterioration of the optical properties emerged for growth conditions away from the ideal Ba to TTIP flux ratio. It was found that a broad optical transition at wavelengths within the near-infrared evolved with increasing TTIP flux, which was attributed to Ba vacancy site formation and TiO$_x$ nanocluster formation in the film.

## 2. Experimental Section

LSAT substrates (lattice parameter 3.8686 Å) with (100) orientation were cleaned by sonication in acetone and isopropyl alcohol followed by a 5-min UV/ozone cleaning in a Boekel Industries UV Clean Model 135500 prior to loading them into the MBE system. BaTiO$_3$ thin films were grown in a DCA M600 hMBE reactor equipped with a Ba thermal effusion cell and a heated metal-organic gas inlet system to supply the metal-organic titanium tetraisopropoxide (TTIP). Additional details of the hMBE setup can be found elsewhere.[38] The TTIP flux was controlled by maintaining a constant gas inlet pressure, $p_{TTIP}$, while the Ba flux was controlled using a PID-controlled effusion cell.[38] LSAT substrates were transferred into the growth reactor and directly heated to a growth temperature of 720 °C, measured by a thermocouple of the substrate heater. Before growth, no surface reconstructions were observed on LSAT substrates by RHEED. The growth conditions were mapped by choosing a fixed Ba flux of $2.50 \times 10^{13}$ cm$^{-2}$s$^{-1}$ with less than 1% drift per hour measured by a quartz crystal microbalance at the sample position. A series of samples were grown using different $p_{TTIP}$ pressures ranging from 56 mTorr to 74 mTorr. BaTiO$_3$ films were grown for 1 h with no additional oxygen, resulting in film thicknesses of about 40 nm.

Films were cooled down in the growth reactor and RHEED images were taken at around 100 °C along the [100] azimuth. Surface morphology measurements were performed using a Bruker Dimension Icon atomic force microscope (AFM) operated in peak force tapping mode. The structural quality of the samples was characterized by high-resolution X-ray diffraction (XRD) scans using a Phillips X'Pert Panalytical MRD Pro by collecting on-axis 2θ-ω scans of the 002 film and substrate peaks and reciprocal space maps in the vicinity of the 103 substrate reflection. High-resolution scanning transmission electron microscopy (STEM) was performed at 300 kV on a dual spherical aberration-corrected FEI Titan G2 60-300 S/TEM. All the STEM images were collected by using a high-angle annular dark field (HAADF) detector with a collection angle of 50-100 mrad. Cross-sectional TEM specimens were prepared using an FEI Helios 660 focused ion beam (FIB) system. A thick protective amorphous carbon layer was deposited over the region of interest then Ga+ ions (30kV then stepped down to 1kV to avoid ion beam damage to the sample surface) were used in the FIB to make the specimen electron transparent for TEM images. In-plane and out-of-plane strain maps, $\varepsilon_{xx}$ and $\varepsilon_{zz}$ were extracted using the Strain++ software based on geometric phase analysis (GPA)[42–44] by taking the substrate as reference. The maps were uploaded in ImageJ[45] and line scans of these maps along the z direction were averaged using a Python script to plot the Strain (%) with respect to the out-of-plane distance. Room temperature ellipsometry spectra in Ψ and Δ were recorded at incident angles of 45°, 55°, 65°, and 75° using a rotating-compensator spectroscopic ellipsometer (J. A. Woollam, M2000) over the spectral range from 375 nm to 1690 nm. The complex refractive index values of n and k were extracted from these spectra using least-squares regression analysis and an unweighted error function to fit a Lorentz oscillator-based model of the dielectric function of a semi-infinite LSAT substrate and $BaTiO_3$ film structure to experimental data. An anisotropic model accounting for the birefringence of the material was constructed by performing Mueller matrix measurements at all angles of incidence measured. An initially isotropic model accounting for only ordinary refractive index, $n_0$, was constructed using data collected at 45°, which was then used to create an identical model for extraordinary refractive index, $n_e$. Oscillator parameter values in $n_e$ were subsequently unconstrained and allowed to converge when fitting data taken at 75°. Both $n_o$ and $n_e$ were then unconstrained and fit parameters were allowed to converge while fitting the entire data set simultaneously in order to obtain the final birefringent model for the indices of refraction. Surface roughness determined from AFM was included in the model and represented by a Bruggeman effective medium approximation of 0.5 void + 0.5 film material fractions.[46]

## 3. Results and Discussions

**Figure 1** shows the RHEED pattern of $BaTiO_3$ films after growth. The highest quality diffraction images with the brightest intensity and narrowest RHEED streaks were found for the film grown at $p_{TTIP}$=60 mTorr. Kikuchi lines were clearly observed and a 4x surface reconstruction appeared along [100], indicating a high crystalline quality of the film and a smooth surface. With decreasing gas inlet pressure $p_{TTIP}$, the RHEED pattern became more diffuse, but Kikuchi lines were still visible. Wider RHEED streaks were found at $p_{TTIP}$=58 mTorr and a surface reconstruction was no longer visible. Further lowering $p_{TTIP}$ down to 56 mTorr, the RHEED image was spotty with very weak diffraction contrast.

For the film grown at $p_{TTIP}$=68 mTorr, the surface reconstructions as well as Kikuchi lines were still visible, although they significantly faded compared to films grown at $p_{TTIP}$=60 mTorr. Further increasing $p_{TTIP}$ resulted in a more pronounced degradation of the RHEED pattern, and Kikuchi lines vanished within an enhanced background intensity. Additionally, intensity modulation along the RHEED diffraction rods emerged for films grown at $p_{TTIP}$=74 mTorr. From these observations, a BaTiO$_3$ growth window cannot be unambiguously determined from the RHEED observation alone on the Ti-rich side. In particular, the gradual transition of RHEED features for films grown at p$_{TTIP}$ pressures higher than 68 mTorr did not allow us to clearly determine when growth conditions changed so that Ti-rich rather than cation stoichiometric films were grown instead. This is in contrast to previous observations for the growth of SrTiO$_3$[47–49] and CaTiO$_3$[50] films by hMBE, where the transition from a self-regulated growth to Ti-rich growth conditions could be unambiguously identified by RHEED. However, the transition from a spotty RHEED diffraction pattern at p$_{TTIP}$=58 mTorr to a streaky pattern, which contained Kikuchi lines and surface reconstructions at p$_{TTIP}$=60 mTorr is very typical and marked the transition from Ba-rich growth conditions to entering condition enabling a self-regulated growth in hMBE.

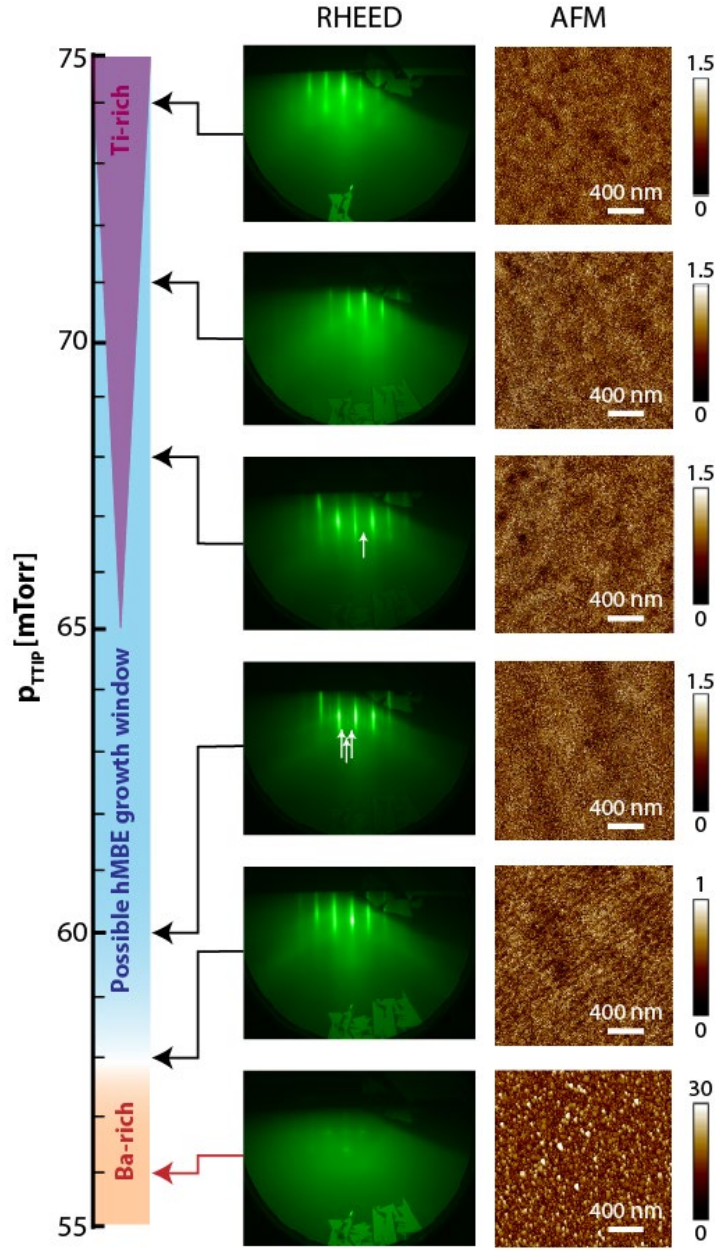

**Figure 1**. (left) Reflection high-energy electron diffraction (RHEED) images taken along the [100] direction and (right) AFM micrographs of BaTiO$_3$ films grown at different p$_{TTIP}$ gas foreline pressures in mTorr (far left) are shown. A transition from the Ba-rich to self-regulated growth conditions was observed in RHEED and AFM, while no clear indication of a transition between self-regulated and Ti-rich growth conditions could be deduced from RHEED and AFM. The surface reconstructions appearing along [100] for the films grown at p$_{TTIP}$=60 and 68 mTorr are indicated with white arrows.

To obtain further insights into the film's stoichiometry AFM images were taken to map the film surface morphology across the stoichiometric series. Films grown under Ba-rich conditions were dominated by an island-like surface morphology. These islands had lateral dimensions on the order of several tens of nanometers and gave rise to substantial surface height modulations with an RMS value of (30±10) Å. BaTiO$_3$ films grown at TTIP pressures between 58 mTorr and 71 mTorr had an atomic terrace morphology with a root means square roughness (RMS) value of 1.5 Å. The number of spatially extended, shallow, and about 1.5 nm-deep holes increased for films grown at higher $p_{TTIP}$ pressures. Again, no clear transition from self-regulated to Ti-rich growth conditions was observed.

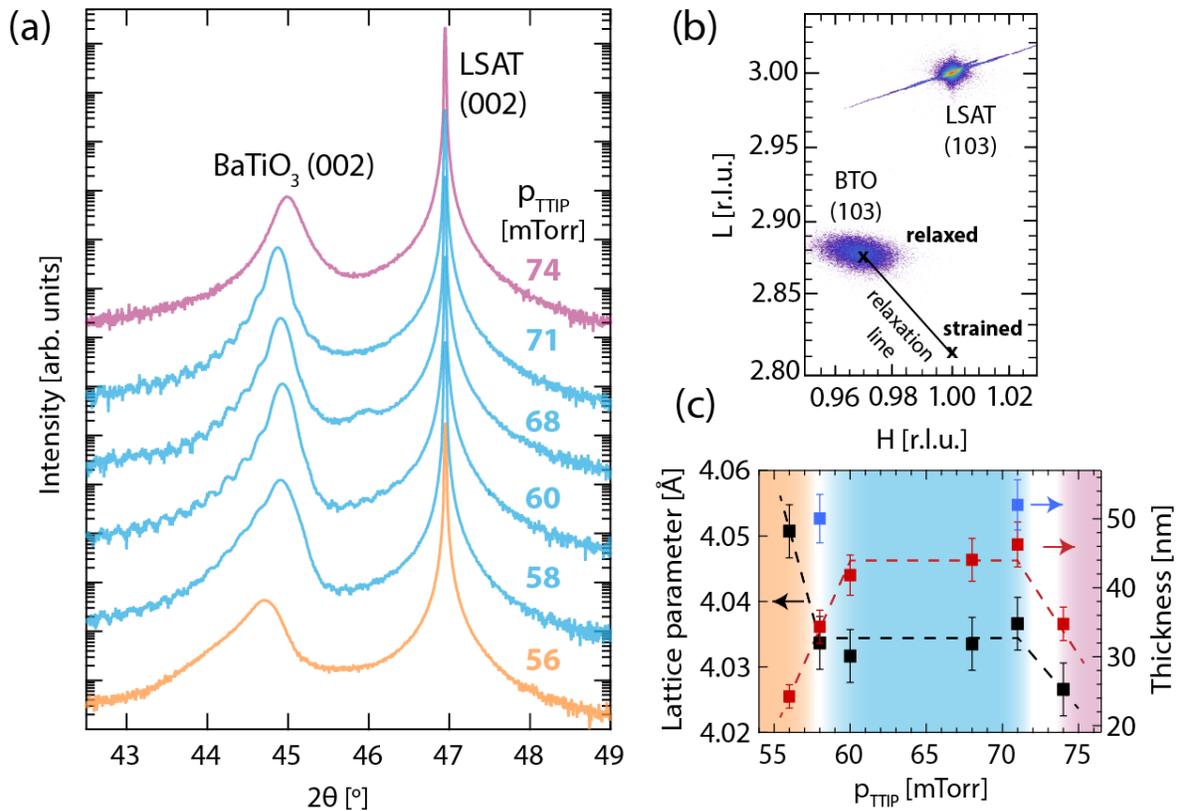

**Figure 2**. (a) High-resolution 2θ-ω X-ray scans around the 002 diffraction peak of LSAT substrate and BaTiO$_3$ films grown at varying $p_{TTIP}$ values. (b) X-ray reciprocal space map around the 103 reflection of LSAT substrate and BaTiO$_3$ grown at $p_{TTIP}$=58 mTorr. (c) Out-of-plane film lattice parameters (black dots) and film thickness (red dots) extracted from (a) using 002 film peak position and width. The dotted lines are a guide to the eyes. Blue dots represent the thicknesses estimated from STEM images.

High-resolution 2θ-ω X-ray scans of the BaTiO$_3$ growth series were taken around the 002-diffraction peak and are shown in **Figure 2(a)**. Films grown under Ba-rich conditions (p$_{TTIP}$=56 mTorr) had a larger out-of-plane lattice parameter, and X-ray film peaks were shifted towards smaller 2θ values. BaTiO$_3$ films grown in the p$_{TTIP}$ range of 58 mTorr to 71 mTorr had a constant out-of-plane lattice parameter and revealed Kiessig fringes, indicating abrupt surfaces and interfaces, which is commonly referred to as an indication of high film quality, while for higher p$_{TTIP}$ values (74 mTorr), the film peak shifted to larger 2θ values. Within the p$_{TTIP}$ range with constant BaTiO$_3$ film lattice parameter, the X-ray peaks had a pronounced asymmetry at 58 mTorr and 60 mTorr, which diminished for films grown at 68 mTorr and 71 mTorr, respectively. No a-domains of BaTiO$_3$ could be detected with RSM, see Supporting Information Figure S1 and S2 for the RSM images of all samples presented in this work.

From the reciprocal space maps (RSM) taken around the 103 substrate and film X-ray reflection shown exemplarily in **Figure 2(b)** for BaTiO$_3$ grown at p$_{TTIP}$=58 mTorr, all films irrespective of growth conditions were found to be nearly completely relaxed (See Supporting Information Figure S1 and S2). This was attributed to the relatively large film thickness and the large epitaxial mismatch imposing a compressive strain of 3.2%. This scenario suggested that the out-of-plane lattice parameter extracted from the XRD scans can be utilized as a direct measure of stoichiometry. **Figure 2(c)** summarizes the out-of-plane lattice parameter and thicknesses of the BaTiO$_3$ films. The data were extracted from the 2θ-ω scans in Figure 2(a) using film peak position and full width at half maximum (FWHM), respectively. The intrinsic lattice parameter of 4.034 Å was found for films grown with p$_{TTIP}$ foreline pressures ranging from 58 to 71 mTorr, which was very close to the expected c-axis lattice parameter of unstrained single crystal BaTiO$_3$ of 4.036 Å.[21] It significantly increased to 4.051 Å for films grown under Ba-rich conditions and was found to be smaller for Ti-rich growth conditions (4.027 Å). Similar trends were found from rocking curve measurements of the 002 BaTiO$_3$ film peak, see Supporting Information Figure S3. While the narrowest rocking curves were found for films grown at p$_{TTIP}$=58 mTorr ($\Delta\omega$=0.070°) and p$_{TTIP}$=60 mTorr ($\Delta\omega$=0.059°), a sizeable increase to about $\Delta\omega$=0.12° was found for films grown under Ba-rich (p$_{TTIP}$=56 mTorr) and Ti-rich (p$_{TTIP}$=74 mTorr) conditions. For BaTiO$_3$ films grown at p$_{TTIP}$=71 mTorr and p$_{TTIP}$=68 mTorr, rocking curve widths of $\Delta\omega$=0.088° and $\Delta\omega$=0.086° were found.

The structural characterization of the stoichiometric series by XRD allowed for refining the MBE growth window edges. X-ray data confirmed that BaTiO$_3$ grown at p$_{TTIP}$=58 mTorr defined the growth window edge towards Ba-rich growth conditions. The transition to Ti-rich growth conditions seemed to occur for p$_{TTIP}$ values between 71 and 74 mTorr, coinciding with a reduced film thickness, reduced out-of-plane lattice parameter, and wider rocking curves.

To directly confirm this conclusion, high-angle annular dark-field (HAADF) scanning transmission electron microscopy (STEM) images were taken for BaTiO$_3$ films grown at p$_{TTIP}$=58 mTorr and p$_{TTIP}$=71 mTorr to ensure cation stoichiometry and the absence of structural defect types of the BaTiO$_3$ films arising from cation nonstoichiometry. Representative HAADF-STEM images of the films taken along the ⟨100⟩ zone axis are shown in **Figure 3**. A highly ordered atomic arrangement was observed for BaTiO$_3$ grown at 58 mTorr resembling the crystalline structure of the perovskite phase, see Figure 3(a). Potential defect formations to accommodate excess Ba in the perovskite structure, namely Ruddlesden-Popper stacking faults, were not found. Although the film showed an ideal perovskite structure, some local areas in the STEM appeared 'blurry', encircled by a white dotted line in Figure 3(a). These features were attributed to local strain fields originating from dislocations formed during film growth to release the epitaxial stress.

To confirm the local nature of the strain fields, in-plane and out-of-plane strain maps, $\varepsilon_{xx}$ and $\varepsilon_{zz}$ (Figure 3(b), 3 (c)) were extracted using the Strain++ software based on geometric phase analysis (GPA)[42–44]. Later, these maps were uploaded in ImageJ[45] and the Strain (%) with respect to the out-of-plane distance was plotted as shown in Figure 3(d). While localized in-plane strain fields were found at the interface and within the BaTiO$_3$ film, not local, but a global strain gradient was found in the out-of-plane strain maps $\varepsilon_{zz}$ (Figure 3(c)). Here, largest strain values were found in the proximity of the BaTiO$_3$/LSAT interface, which then decreased with increasing distance away from the interface and finally remained constant at a distance about 20 nm away from it, see Figure 3(d). The strain gradient present in the film grown at 58 mTorr explained the pronounced X-ray peak asymmetry. The larger strain state of BaTiO$_3$ closer to the interface resulted in an expansion of the out-of-plane lattice parameter, which shifted the weight of the X-ray peak towards smaller 2θ values. It is concluded that a strain gradient is also present in the film grown at p$_{TTIP}$=60 mTorr, but not in films grown at higher p$_{TTIP}$. The presence of a strain gradient in the film is also evident from reciprocal space maps shown in Supporting Information Figure S1. Here, the film peak's largest full width at half maximum was tilted towards the relaxation line for BaTiO$_3$ with an in-built strain gradient, while for films without strain gradient the largest full width at half maximum was aligned with the in-plane direction. The asymmetry of the 002 X-ray film peak of BaTiO$_3$ grown at p$_{TTIP}$=58 mTorr also caused a widening in the FWHM value and resulted in a relatively lower film thickness extracted from the X-ray film peak width (estimated to be about 35 nm) compared to the actual film thickness of about 50 nm determined by STEM, see Figure 2(c). It is obvious from the low-magnification STEM images that the thickness of the sample grown at p$_{TTIP}$=58 mTorr, in particular, is about 50 nm suggesting that the FWHM estimations were inaccurate for the samples with asymmetric XRD peaks. While for the sample grown at 71 mTorr which was shown to have the most symmetric XRD peak, the calculated thickness from the X-ray peak width (~46 nm) was somewhat closer to the actual value measured by STEM (52 nm) (Supporting Information Figure S4). The thickness values of these two samples determined by STEM are shown as blue data points in Figure 2(c).

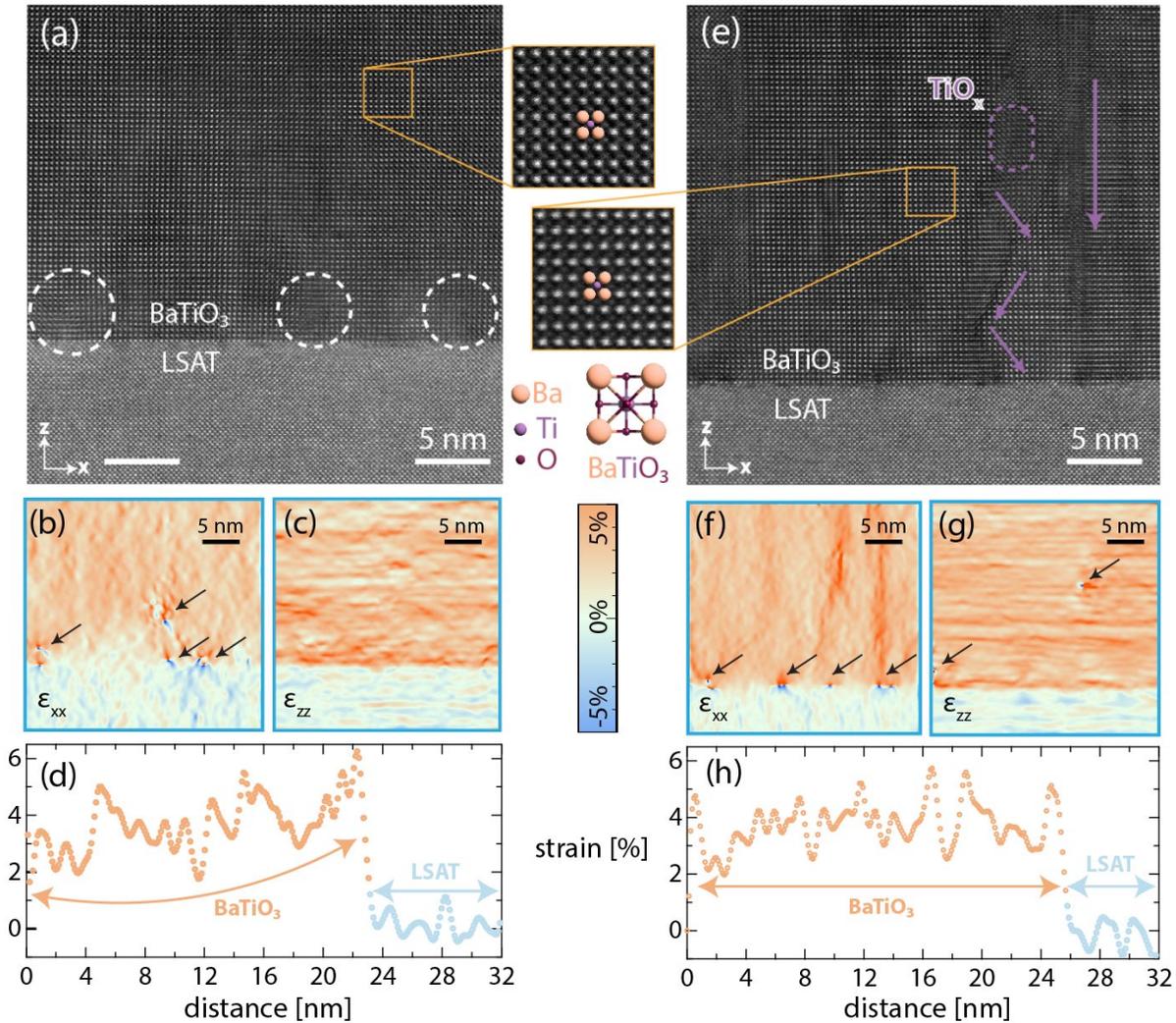

**Figure 3**. High-angle annular dark-field scanning transmission electron microscopy (HAADF-STEM) images of BaTiO$_3$ thin films grown at (a) p$_{TTIP}$=58 mTorr and (e) p$_{TTIP}$=71 mTorr. White dotted circles in (a) highlighting 'blurry' sections arising from localized strain fields. Purple dotted circles and arrows in (e) emphasize nanosized crystalline inclusions and interfaces formed with the BaTiO$_3$ perovskite phase. Strain maps extracted from STEM image in (a) using the Strain++ software based on geometric phase analysis (GPA)[42–44] for (b) in-plane $\varepsilon_{xx}$, and (c) and out-of-plane $\varepsilon_{zz}$ maps of BaTiO$_3$ grown at p$_{TTIP}$=58 mTorr. Local strain fields from dislocations are highlighted with black arrows. (d) Average line scans extracted from out-of-plane strain maps $\varepsilon_{zz}$ shown in (c) revealing a strain gradient in the vicinity of the BaTiO$_3$/LSAT interface. Strain maps extracted from STEM image in (e) for (f) in-plane $\varepsilon_{xx}$, and (g) out-of-plane $\varepsilon_{zz}$ maps of BaTiO$_3$ grown at p$_{TTIP}$=71 mTorr. Local strain fields from dislocations are highlighted with black arrows. (h) Average line scans extracted from out-of-plane strain maps $\varepsilon_{zz}$ shown in (g) with no a strain gradient in BaTiO$_3$.

The situation is very different for BaTiO$_3$ grown at p$_{TTIP}$=71 mTorr, see Figure 3(e)-(h) where the same type of strain analysis was applied as well. While the majority of the film was in its perovskite phase, nanosized inclusion of an additional phase was found as well. The absence of very bright atomic columns was indicative that this undesired phase did not contain any Ba and was most likely anatase TiO$_2$. The presence of these nanoscale inclusions affected the overall strain state of the film. Strong and localized strain fields were still present in $\varepsilon_{xx}$, see Figure 3(f). In addition, larger in-plane strain modulations that coincided with the presence of the additional phase were found, while the overall strain gradient in the out-of-plane direction (see $\varepsilon_{zz}$ maps shown in Figure 3(g) and the line plot in Figure 3(h)) was not detected. A different strain accommodation mechanism enabled by the presence of nanosized Ba-deficient crystalline phases ultimately rendered the out-of-plane film lattice parameter an irrelevant metric to determine the film's cation stoichiometry. Therefore, the BaTiO$_3$ growth window is expected to be smaller than anticipated from the X-ray analysis. If the absence of a strain gradient is coincident with the formation of an undesirable Ba-deficient phase, it is to be expected that also the BaTiO$_3$ film grown at p$_{TTIP}$= 68mTorr is not cation stoichiometric. This suggests that the p$_{TTIP}$ foreline pressure, for which cation-stoichiometric BaTiO$_3$ can be achieved, was in the range of p$_{TTIP}$= 58mTorr to p$_{TTIP}$= 60mTorr, at best in the range p$_{TTIP}$= 58mTorr to p$_{TTIP}$< 68mTorr. Expressing the growth window edges as $p_{TTIP}^{min}$ and $p_{TTIP}^{max}$ for these two scenarios, a relative growth window width $\Delta$ defined as

$$\Delta = \frac{p_{TTIP}^{max} - p_{TTIP}^{min}}{p_{TTIP}^{max} + p_{TTIP}^{min}}$$

yielded Δ=0.02 and Δ=0.08, respectively. This is somewhat smaller compared to Δ=0.10 estimated for partially relaxed BaTiO$_3$ grown on SrTiO$_3$ at 950 °C,[8] and considerably smaller compared to the relative growth window widths of Δ=0.20 and Δ=0.23 determined for BaTiO$_3$ grown coherently strained on GdScO$_3$(110) at 770 °C and 870 °C, respectively. [9] The discrepancy was attributed to the lower growth temperature of BaTiO$_3$ films here, which typically shrinks the growth window width in hMBE.

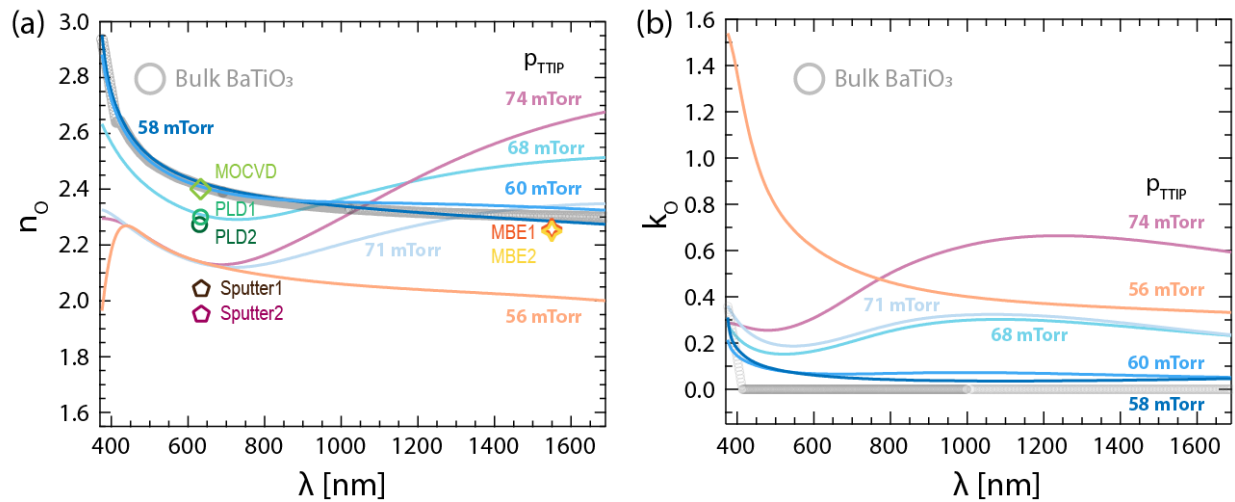

**Figure 4**. (a) Ordinary refractive index n$_o$ and (b) ordinary extinction coefficients, k$_o$ of BaTiO$_3$ thin films grown at p$_{TTIP}$ pressures ranging from 56 mTorr to 74 mTorr. The data extracted from spectroscopic ellipsometry is plotted together with bulk BaTiO$_3$ (gray circles) and BaTiO$_3$ films grown by other techniques that were taken from the references.[4,25,28,33,51]

To further gain insight into the film's stoichiometry and to utilize a physical property as a stoichiometry measure, the optical properties of BaTiO$_3$ films were determined by spectroscopic ellipsometry. **Figure 4** shows the ordinary refraction indices and extinction coefficients of BaTiO$_3$ films grown by hMBE in comparison to bulk BaTiO$_3$.[51] Due to the optical anisotropy of BaTiO$_3$, both ordinary and extraordinary refractive indices and extinction coefficients were extracted, and the extraordinary responses are presented in Supporting Information Figure S5. Since the band gap of BaTiO$_3$ is about 3.2 eV,[51] a large absorption peak and thus a large extinction coefficient were present at 387 nm. Consequently, the index of refraction had a maximum value and continuously decreased with increasing wavelength, while the optical loss was expected to be near zero in the measured spectral range. The stoichiometry of BaTiO$_3$ has a tremendous effect on both, the index of refraction ($n_1$, $n_2$) and optical loss (extinction coefficients, $k_1$ and $k_2$), see Table 1. Films grown at $p_{TTIP}$=58 mTorr and $p_{TTIP}$=60 mTorr had refractive indices that ideally mapped onto bulk BaTiO$_3$ data throughout the entire spectral range. Deviations from ideal stoichiometry irrespective of its type, i.e. either for Ba-rich or Ti-rich BaTiO$_3$ films, resulted in a reduced refractive index, see Figure 4(a). For Ba-rich films grown at $p_{TTIP}$=56 mTorr and Ti-rich films grown at $p_{TTIP}$=71 mTorr and $p_{TTIP}$=74 mTorr, a refractive index of 2.15 was found at a wavelength of 600 nm, about 90% of the value measured for bulk single crystals (2.41). Therefore, the refractive index at a wavelength around 600 nm was found to be an effective metric to indirectly determine the degree of nonstoichiometry in BaTiO$_3$. A direct comparison with the literature revealed that the refractive index data of stoichiometric BaTiO$_3$ was rarely reported for films (MOCVD-grown films: $n_1$=2.40[28]). BaTiO$_3$ films grown by PLD already had a sizeable reduction down to $n_1$=2.32[4] and $n_1$=2.27[33] at 630nm, while BaTiO$_3$ films deposited by sputtering showed even lower values of $n_1$=2.04 and $n_1$=1.96.[25]

**Table 1**. Ordinary refractive indices, $n_1$ and $n_2$, and ordinary extinction coefficients, $k_1$ and $k_2$, of all samples ($p_{TTIP}$= 56-74 mTorr) studied in this work listed together with bulk $BaTiO_3$ and $BaTiO_3$ grown by different growth techniques (MOCVD, PLD, MBE) and at wavelengths 632-635 nm and 1550 nm.

|  | λ (nm) | $n_1$ | $k_1$ |  | λ (nm) | $n_2$ | $k_2$ |
|---|---|---|---|---|---|---|---|
| Bulk BTO1[51] | 632 | 2.41 | 0 | Bulk BTO1[51] | 1550 | 2.30 | 0 |
| Bulk BTO2[52] | 633 | 2.41 | 0.0002 | Bulk BTO2[52] | - | - | - |
| 58 mTorr | 632 | 2.42 | 0.056 | 58 mTorr | 1550 | 2.28 | 0.042 |
| 60 mTorr | 632 | 2.41 | 0.066 | 60 mTorr | 1550 | 2.33 | 0.056 |
| 68 mTorr | 632 | 2.31 | 0.176 | 68 mTorr | 1550 | 2.5 | 0.251 |
| 71 mTorr | 632 | 2.13 | 0.204 | 71 mTorr | 1550 | 2.34 | 0.260 |
| 74 mTorr | 632 | 2.14 | 0.336 | 74 mTorr | 1550 | 2.63 | 0.623 |
| 56 mTorr | 632 | 2.14 | 0.577 | 56 mTorr | 1550 | 2.01 | 0.341 |
| MOCVD[28] | 633 | 2.4 | - | MBE1[18] | 1550 | 2.27 | - |
| PLD1[4] | 633 | 2.32 | - | MBE2[16] | 1550 | 2.26 | - |
| PLD2[33] | 630 | 2.27 | - |  | - | - | - |
| Sputter1[25] | 635 | 2.04 | - |  | - | - | - |
| Sputter2[25] | 635 | 1.96 | - |  | - | - | - |

In the visible range at wavelengths between 400 nm and 800 nm, the index of refraction was continuously reduced with an increasing degree of nonstoichiometry for $BaTiO_3$ films studied in this work. On the contrary, in the near-infrared region, the change of refraction index with nonstoichiometry was found to depend on the type of nonstoichiometry. Excess Ba in the film ($p_{TTIP}$=56 mTorr) resulted in a decrease in the refractive index, while films grown under Ti-rich conditions revealed an upturn in the IR region, even exceeding the values of stoichiometric $BaTiO_3$. For the technologically relevant wavelength of 1550 nm, a value of $n_2$=2.30 was found for bulk $BaTiO_3$ while films grown by hMBE at $p_{TTIP}$=58 mTorr and $p_{TTIP}$=60 mTorr had an index of refraction of $n_2$=2.28 and $n_2$=2.33, respectively. Ba-rich films had values below $n_2$=2.1, while Ti-rich $BaTiO_3$ had $n_2$ values higher than stoichiometric $BaTiO_3$ at 1550 nm.

An increase in the extinction coefficients $k_o$ throughout the entire spectrum was found for nonstoichiometric $BaTiO_3$ films. $BaTiO_3$ films grown at $p_{TTIP}$=58 mTorr and $p_{TTIP}$=60 mTorr had $k_o$ values very close to bulk $BaTiO_3$, see Figure 4(b). It is noted that in both films a sizeable absorption was still present in the wavelength interval between 400 nm to 600 nm, i.e. in the vicinity of the interband transition at 387 nm, potentially indicating the presence of Urbach tails[53,54] arising from disorder or defects, such as dislocations or point defects present in the films. With an increasing level of excess Ti in the film, a wide $k_o$ peak emerged with a maximum of around 1 μm, which became more pronounced and shifted towards larger wavelengths with an increasing level of nonstoichiometry. Again, the optical loss can be utilized as a metric for nonstoichiometry, and although it is much more straightforward to interpret and link to specific defects formed, the lack of $BaTiO_3$ film data in the literature makes it somewhat less relevant.

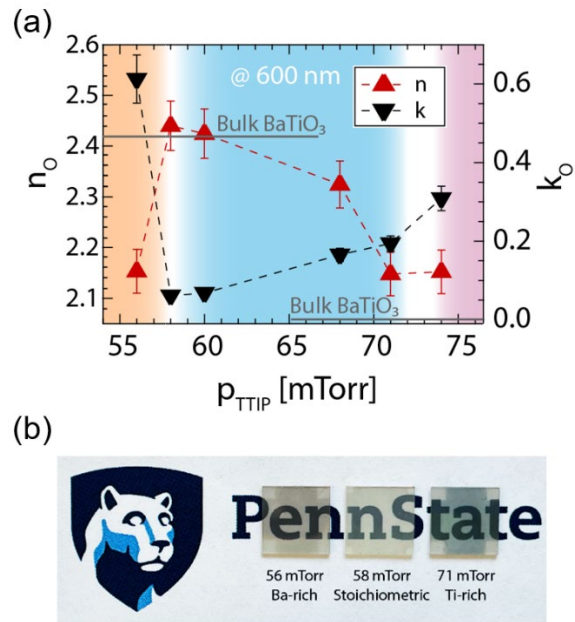

**Figure 5**. (a) Ordinary refractive index $n_o$ and extinction coefficient $k_o$ as a function of gas inlet pressure $p_{TTIP}$ at a wavelength of 600 nm. (b) A picture taken from $BaTiO_3$ thin film grown on LSAT ($p_{TTIP}$ = 58 mTorr) and its high optical transparency compared to films grown under Ba-rich ($p_{TTIP}$ = 56 mTorr) and a Ti-rich ($p_{TTIP}$ = 71 mTorr) growth conditions.

**Figure 5(a)** summarizes the results of the optical property measurements with changes in the growth conditions. The monotonous behavior of the index of refraction and optical loss at 600 nm makes them an ideal measure of nonstoichiometry in $BaTiO_3$ and helps to identify the growth window in which the stoichiometry of $BaTiO_3$ grown by hMBE was self-regulated. While $n_0$ of $BaTiO_3$ films grown at $p_{TTIP}$=58 and 60 mTorr matched the values of bulk $BaTiO_3$ within the error of the experiment, the smallest loss was found for these films as well. Deviation from these growth conditions resulted in an increase in optical loss and a reduction of refractive index. This suggests that a growth window, in particular, the edge towards Ti-rich growth conditions cannot be unambiguously determined by AFM, RHEED, and X-ray, but rather by STEM and optical properties. The latter is much more straightforward suggesting that spectroscopic ellipsometry constitutes a better metric for the intrinsic properties of the samples. **Figure 5(b)** demonstrates the high optical transparency of a stoichiometric $BaTiO_3$ film grown at $p_{TTIP} = 58$ mTorr compared to the relatively low optical transparency of films grown under Ba-rich ($p_{TTIP} = 56$ mTorr) or a Ti-rich ($p_{TTIP} = 71$ mTorr) conditions.

4. **Conclusion**

In summary, $BaTiO_3$ thin films were grown by hMBE and found to show bulk-like optical properties in a narrow growth window. Films grown under deviating conditions revealed inferior optical properties, namely a lower refractive index in the visible spectrum and a larger absorption in the visible and near IR. These trends were found to be a robust indicator for the degree and the type of cation non-stoichiometry in the films, adding another, yet underemployed method towards quantifying nonstoichiometry in $BaTiO_3$ films. Conventionally employed characterization techniques, such as RHEED, AFM, and XRD allowed identifying the growth window edge towards Ba-rich growth conditions, while the determination of Ti-rich conditions was found to be notoriously challenging. The ability to grow transparent $BaTiO_3$ with superior optical properties opens opportunities for increasing the scalability and potentially enables increased output efficiency for Pockels-based electro-optic modulator devices in advanced silicon photonic platforms. While the growth of $BaTiO_3$ on LSAT substrates is not necessarily representative of the full potential the $BaTiO_3$ thin film material can achieve, this work provides a solid basis for the hMBE growth of $BaTiO_3$ and emphasizes the potential that $BaTiO_3$ can unleash to reach its full potential also in the form of its thin film.

**Data availability**

The data that support the findings of this study are available from the corresponding authors, B. F. Y. and R. E.-H., upon reasonable request.

**Competing Interests**

The authors declare no conflict of interest.

**Acknowledgments**

This work is supported by the US Department of Energy, Office of Science, Basic Energy Sciences, under Award Number DE-SC0020145 as part of the Computational Materials Sciences Program. B. F. Y. would also like to thank Saiphaneendra Bachu for the fruitful discussions on TEM imaging and strain maps, and also Irene Ayuso Pérez for her support for the Python script to generate the strain plots.

Supporting Information

**Stoichiometric Control and Optical Properties of BaTiO$_3$ Thin Films Grown by Hybrid MBE**

*Benazir Fazlioglu Yalcin[*], Albert Suceava, Tatiana Kuznetsova, Ke Wang, Venkatraman Gopalan, and Roman Engel-Herbert[*]*

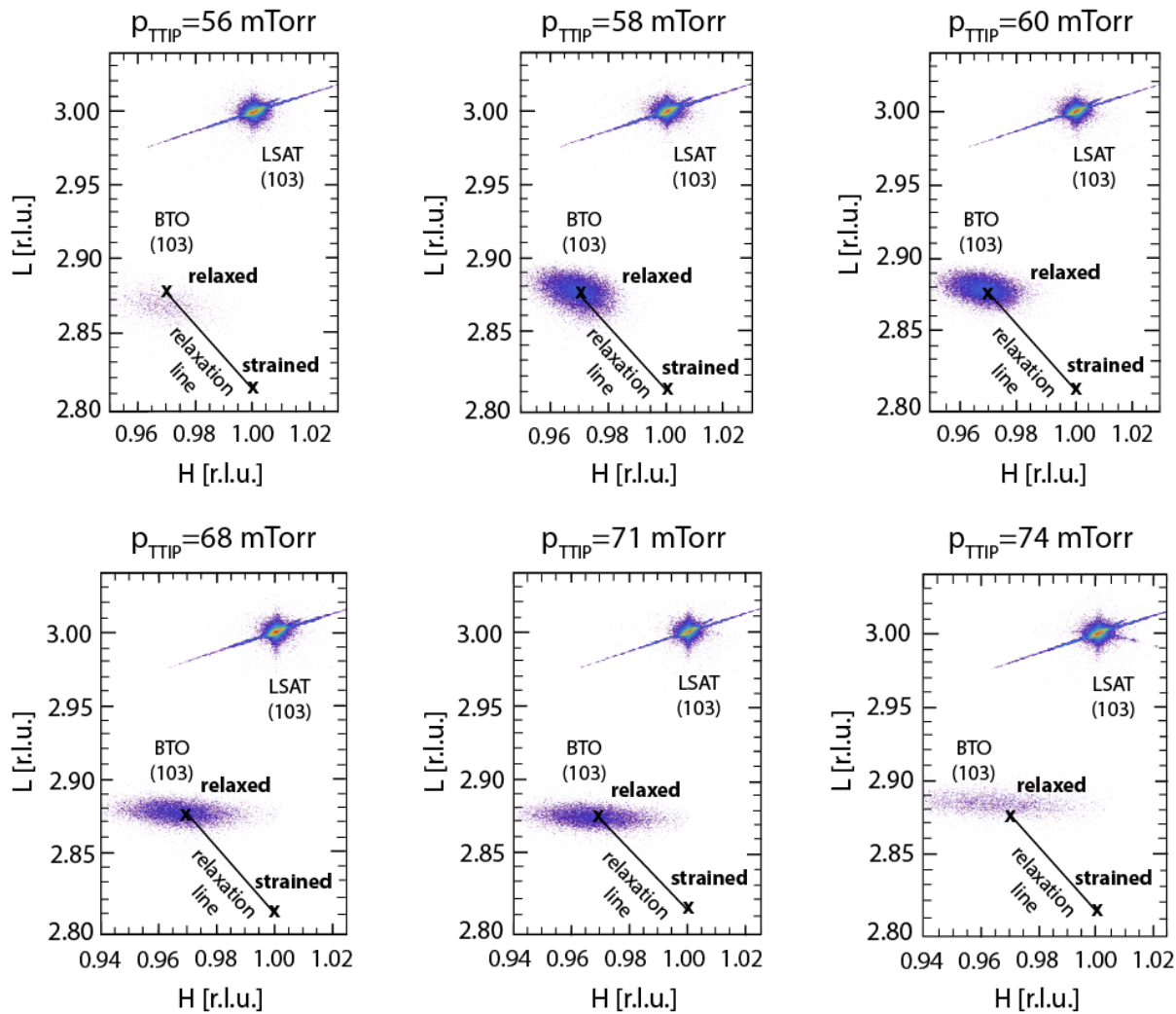

**Figure S1.** X-ray reciprocal space maps of BaTiO$_3$ thin films grown by hMBE on LSAT with varying p$_{TTIP}$ gas inlet pressures ranging from 56 mTorr to 74 mTorr. The maps were taken around the 103 diffraction peaks and the maps are given in reciprocal lattice units (r.l.u.) of the LSAT substrates with lattice parameter 3.868 Å. All BaTiO$_3$ thin films irrespective of growth conditions were relaxed on LSAT; the film reflection of the samples grown at p$_{TTIP}$ 58 and 60 mTorr were tilted towards the relaxation line, attributed to the presence of a strain gradient present in the proximity of the film/substrate interface.

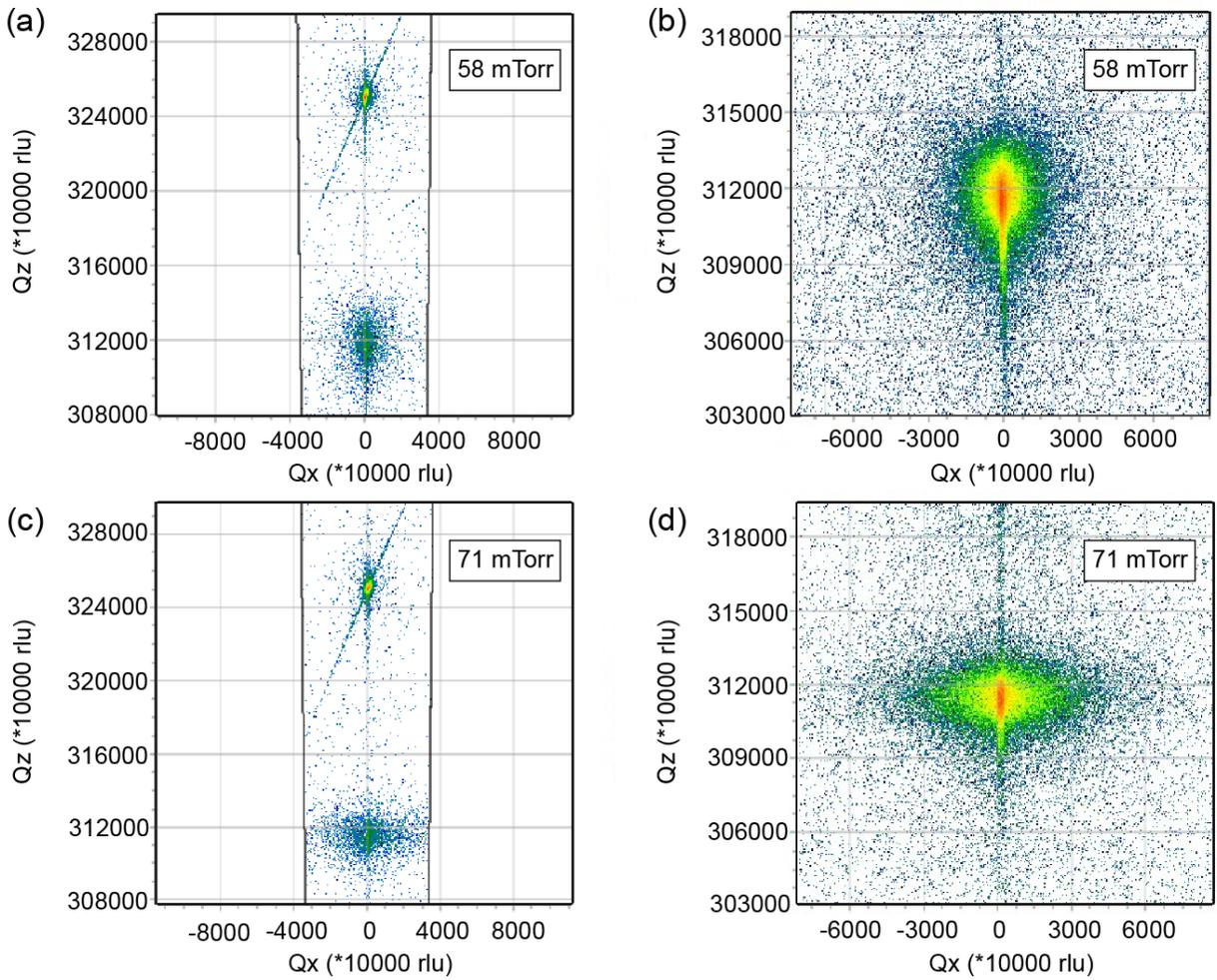

**Figure S2.** X-ray reciprocal space maps (2θ-ω) around the 002 diffraction peak of the BaTiO$_3$ films grown at (a) 58 mTorr and (c) 71 mTorr. X-ray reciprocal space maps (ω) around the 002-diffraction peak of the BaTiO$_3$ films grown at (b) 58 mTorr and (d) 71 mTorr.

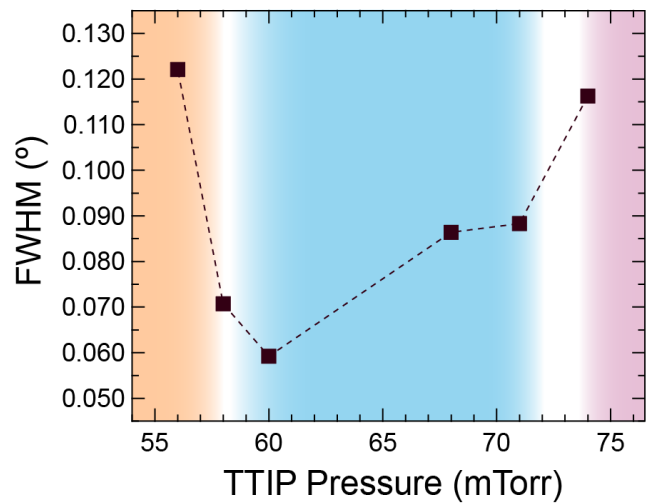

**Figure S3.** Rocking curve width of the 002 BaTiO$_3$ film peak for the stoichiometric series of BaTiO$_3$ films grown by hMBE with $p_{TTIP}$ gas inlet pressures ranging from 56 mTorr to 74 mTorr.

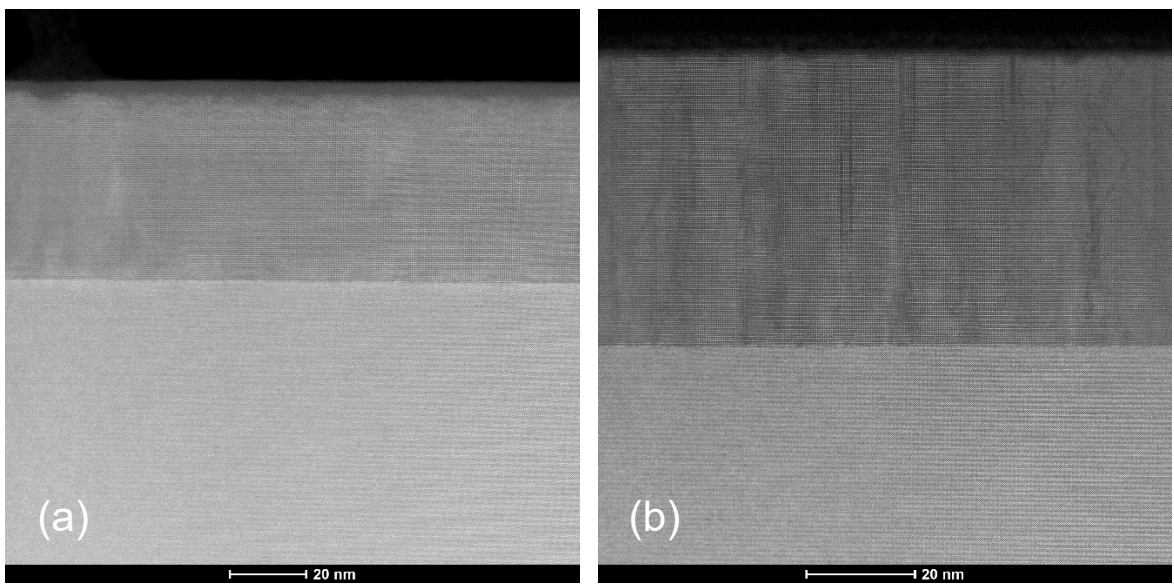

**Figure S4.** High-angle annular dark-field scanning transmission electron microscopy (HAADF-STEM) images of BaTiO$_3$ thin films grown at (a) p$_{TTIP}$=58 mTorr, and (b) p$_{TTIP}$=71 mTorr. The film thicknesses are measured to be 50 nm and 52 nm, respectively.

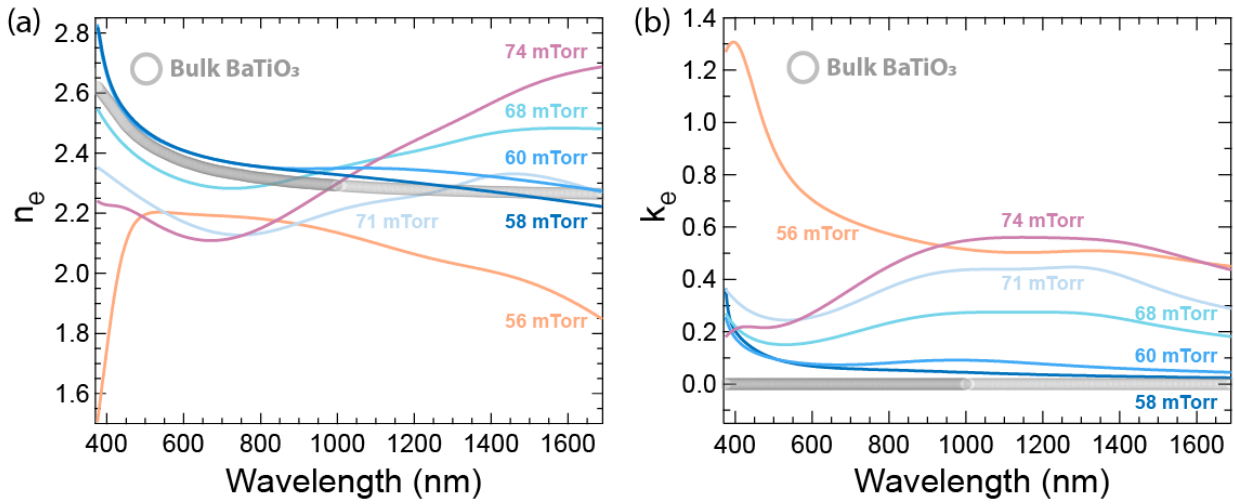

**Figure S5.** (a) Extraordinary refractive index $n_e$ and (b) extraordinary extinction coefficients $k_e$ of BaTiO$_3$ thin films grown at $p_{TTIP}$ ranging from 56 mTorr to 74 mTorr. All data were plotted together with bulk BaTiO$_3$ values (gray circles) taken from Ref. [1].